# Controlled Fabrication of Native Ultra-Thin Amorphous Gallium Oxide from 2D Gallium Sulfide for Emerging Electronic Applications

AbdulAziz AlMutairi[1,2], Aferdita Xhameni[2,3], Xuyun Guo[4,5,6], Irina Chircă[1], Valeria Nicolosi[4,5,6], Stephan Hofmann[1], Antonio Lombardo[2,3]

[1]Department of Engineering, University of Cambridge, United Kingdom

[2]London Centre for Nanotechnology, 19 Gordon St, London, WC1H 0AH, United Kingdom

[3]Department of Electronic & Electrical Engineering, Malet Place, University College London, WC1E 7JE, United Kingdom

[4]Advanced Materials and BioEngineering Research (AMBER), Centre at Trinity College Dublin and the Royal College of Surgeons in Ireland, Dublin, 2, Ireland

[5]Trinity Centre for Biomedical Engineering, Trinity College Dublin, Dublin, 2, Ireland

[6]School of Chemistry, Trinity College Dublin, Dublin, 2, Ireland

## Abstract

Oxidation of two-dimensional (2D) layered materials has proven advantageous in creating oxide/2D material heterostructures, opening the door for a new paradigm of low-power electronic devices. Gallium (II) sulfide (β-GaS), a hexagonal phase group III monochalcogenide, is a wide bandgap semiconductor with a bandgap exceeding 3 eV in single and few layer form. Its oxide, gallium oxide ($Ga_2O_3$), combines large bandgap (4.4-5.3 eV) with high dielectric constant (~10). Despite the technological potential of both materials, controlled oxidation of atomically-thin β-GaS remains under-explored.  This study focuses into the controlled oxidation of β-GaS using oxygen plasma treatment, achieving ultrathin native oxide ($GaS_xO_y$, ~4 nm) and $GaS_xO_y$/GaS heterostructures where the GaS layer beneath remains intact. By integrating such structures between metal electrodes and applying electric stresses as voltage ramps or pulses, we investigate their use for resistive random-access memory (ReRAM). The ultrathin nature of the produced oxide enables low operation power with energy use as low as 0.22 nJ per operation while maintaining endurance and retention of 350 cycles and $10^4$ s, respectively.  These results show the significant potential of the oxidation-based $GaS_xO_y$/GaS heterostructure for electronic applications and, in particular, low-power memory devices.

## Introduction

Gallium oxide ($Ga_2O_3$), a wide bandgap semiconductor, has emerged as a material of significant interest to the semiconductor industry owing to its excellent chemical and electrical properties. Its wide bandgap of 4.4 to 5.3 eV[1-5], in addition to its thermal and chemical stability, made $Ga_2O_3$ applications in electronics very versatile. Notably, β-$Ga_2O_3$ boasts a high breakdown field strength of up to 8 MVcm$^{-1}$ [6], a fact that fuelled the investigation into its applications in high-power electronics. Beyond conventional electronics, $Ga_2O_3$ has demonstrated tremendous potential in resistive switching (RS) applications due to its high intrinsic resistance and sensitivity to oxygen content[7-9]. The latter is more predominantly observed in amorphous $Ga_2O_3$, and low-power uni- and bipolar switching has been achieved in $Ga_2O_3$ memristors[10-16].

Nevertheless, the fabrication of ultra-thin, high-quality film of $Ga_2O_3$ remains a challenge. To this end, several techniques have been explored, such as the mechanical exfoliation of large $Ga_2O_3$ crystals and the squeezing of liquid Gallium metal[15-17]. However, the previously reported techniques, while promising, still face many challenges, such as the potential for scalability. A rarely explored scalable approach to obtain ultra-thin $Ga_2O_3$ is through the oxidation of a 2D layered material containing Ga metal[18]. Oxidation of 2D materials has been shown to produce ultra-thin oxide with high dielectric constant and clean interface in materials such as $HfS_2$[19,20], $HfSe_2$[21], $TaS_2$[22], $ZrSe_2$[21], and $Bi_2O_2Se$[23]. Oxidation techniques spanning thermal[21-23], oxygen plasma[19,20], and photo-oxidation[24] have been explored, with Plasma oxidation standing out for its ability to produce uniform dielectric layers. Furthermore, using a standard resist mask allows precise control of the location of oxidation, offering enhanced fabrication flexibility and tailored device functionality.

Group III monochalcogenides layered materials in X-M-M-X (M = Ga, In and X = S, Se, Te) form present a promising platform for electronic device applications[25-28]. More importantly, it has been reported that a good number of Ga-based group III monochalcogenides form a stable native $Ga_2O_3$[29,30]. One of the members of the monochalcogenides group that is generally neglected is hexagonal-layered gallium (II) sulfide (GaS), commonly referred to as β-GaS. A wide band-gap 2D material, GaS, has found its footing in applications such as optoelectronics[25,26,31]. In line with other layered monochalcogenides, GaS can be oxidised despite not being as rapid as GaSe or InSe[18]. However, reports on the oxidation mechanism of GaS are scarce and primarily focus on incidental environmental exposure rather than controlled oxidative processes[18].

This study investigates gallium oxide ($GaS_xO_y$) and $GaS_xO_y$/GaS structures produced by oxidising GaS via plasma oxidation under carefully controlled conditions and their use in resistive memories. Plasma oxidation of GaS resulted in ultra-thin amorphous $GaS_xO_y$ while keeping the GaS layers beneath intact. Resistive-random-access memory (ReRAM) based on $GaS_xO_y$/GaS heterostructure exhibited low switching voltages and respectable endurance and retention.

## Results and Discussion:

Figures 1 (a-c) show a schematic of the GaS oxidation process using RF oxygen plasma at room temperature. Multilayer GaS flakes are mechanically exfoliated (as described in the methods section) on a SiO$_2$/Si substrate. The flakes are then exposed to 10 W of O$_2$ plasma at 10 sccm and 20 mTorr, which alters the top layers of the flake chemically by converting them into a thin oxide layer. The chemical alteration can be deduced optically by observing a colour change in the flake post-oxidation in which the colour appears more translucent as the thickness of the oxide increases (Fig1 (d) and (e)). The final product could be partially or fully oxidised, depending on the oxidation conditions, time, and the initial GaS flake thickness. Figure 1 (f) shows the Raman spectra of three GaS flakes at different oxidation levels, which are pristine, partially oxidised, and fully oxidised in addition to the spectrum of the SiO$_2$/Si substrate for reference. While both pristine and partially oxidised GaS show the characteristic $A^1_{1g}$, $A^2_{1g}$, and $E^1_{2g}$ peaks, fully oxidised flake shows no characteristic peaks of GaS, as shown in Fig 1 (f).

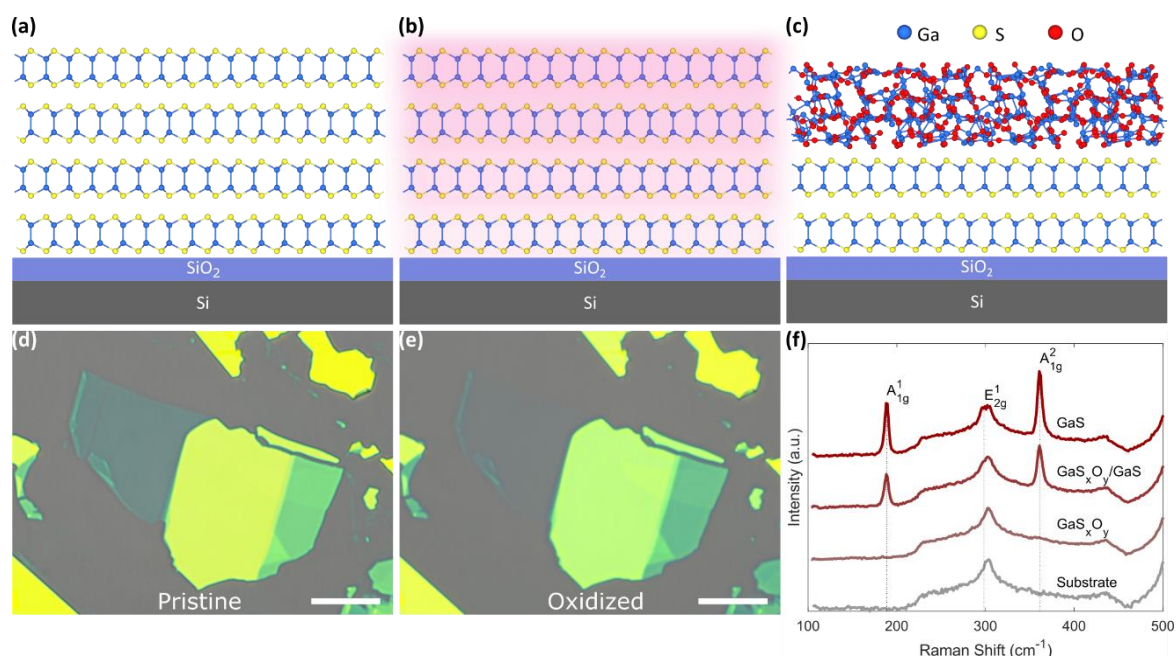

Figure 1: Schematic representation of the oxidation process in GaS. (a) Exfoliation of the GaS flake onto a SiO$_2$/Si substrate. (b) Oxygen plasma exposure of the exfoliated GaS flake. (c) Conversion of the topmost layers into GaS$_x$O$_y$. (d) and (e) shows optical microscope images of a pristine and an oxidised GaS flake on SiO$_2$/Si substrate with on 285 nm SiO$_2$. The sample was oxidised at 10 W of O$_2$ plasma at 10 sccm and 20 mTorr for 15 mins and show a clear colour change post-oxidation (scalebar 10 μm). (f) Raman spectra of three different GaS flakes at different oxidation levels. Pristine GaS and partially oxidised GaS$_x$O$_y$/GaS show the characteristic $A^1_{1g}$, $A^2_{1g}$, and $E^1_{2g}$ peaks, whereas fully oxidised GaS$_x$O$_y$ shows no characteristic peaks. The spectrum of SiO2/Si substrate in shown in grey.

To further characterise GaS and GaS$_x$O$_y$, spectroscopic imaging ellipsometry (SIE) was utilised as described in the methods section. To this end, A ~10.3 nm pristine GaS flake (kept in an oxygen-free environment during measurement) and a fully oxidised ~2.4 nm GaS$_x$O$_y$ (both exfoliated on SiO$_2$/Si substrate) were used for the measurements. The Tauc-Lorentz oscillator model was employed to fit the GaS data [32], whereas the Cauchy dispersion model was utilised for the $GaS_xO_y$.

Figure 2(a) shows the Tauc-Lorentz oscillator model fit of the ψ and Δ of the GaS. From the model, the bandgap was estimated to be 2.44 ± 0.15 eV, which is in good agreement with previous reports [31,33]. Due to the limited energy range used for this measurement, no interband transition peaks were observed in the dielectric response (Fig 2(b)). The Cauchy dispersion, shown in Fig. 2(c), fits well to the transparent $GaS_xO_y$. The refractive index (n), shown in Fig. 2(d), differs from the typical refractive index reported for amorphous $Ga_2O_3$. For example, at 600 nm wavelength, fully oxidised $GaS_xO_y$ exhibit n ~1.6, whereas the n value reported in the literature for amorphous $Ga_2O_3$ sits in the range of 1.75-1.9[34-36]. This observation suggests that the oxidised GaS is of lower density, leading to a lower refractive index[37].

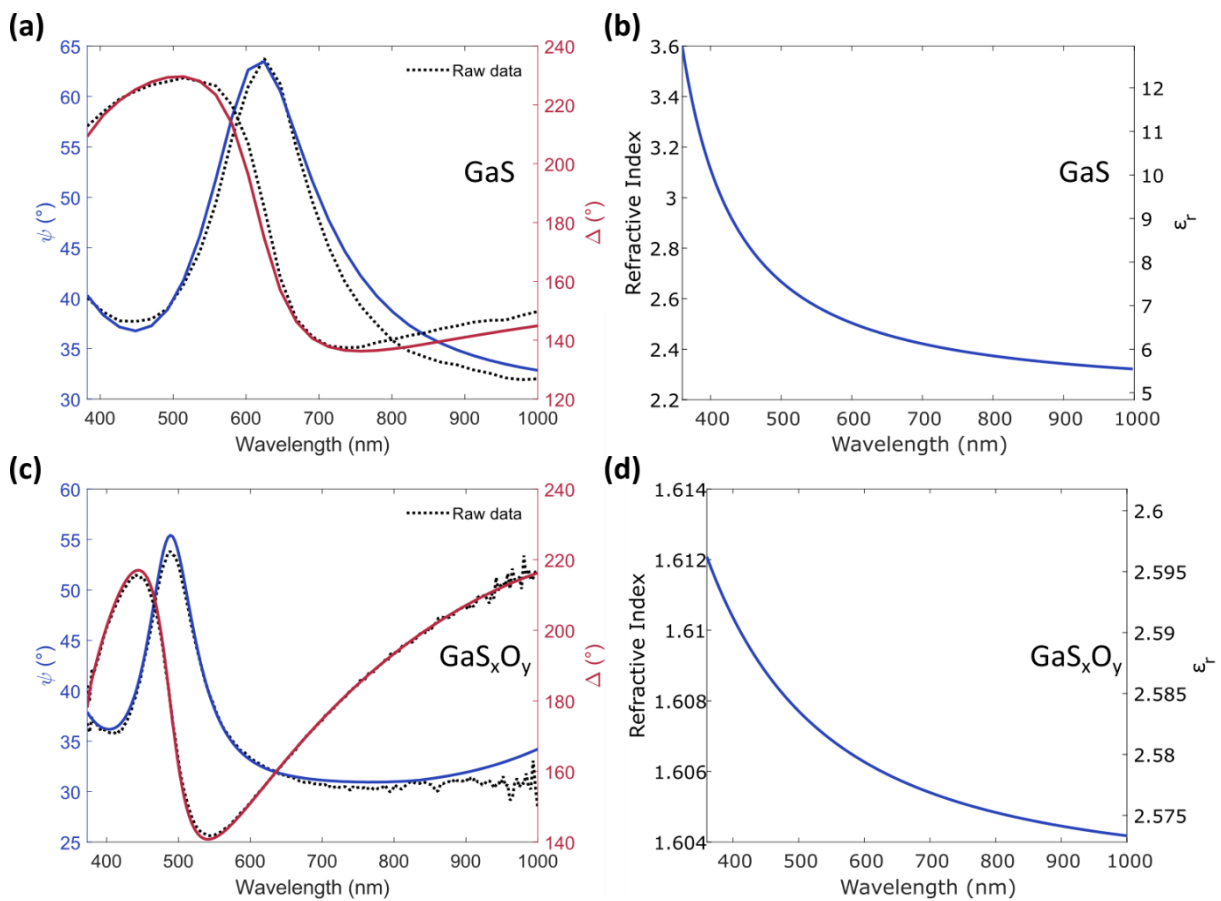

**Figure 2: Spectroscopic imaging ellipsometry (SIE) of GaS and $GaS_xO_y$.** (a) The ψ and Δ spectra fitted using the Tauc-Lorentz oscillator model (A = 94.3 ± 17.2 eV, $E_0$ = 4.133 ± 0.004 eV, $E_g$ = 2.44 ± 0.15 eV, and Γ = 0.05 ± 1 eV). (b) is the corresponding refractive index and the real part of the dielectric response. (c) The ψ and Δ spectra of $GaS_xO_y$ fitted using the Cauchy dispersion model (A = 1.603 ± 0.005, B = 1177 ± 1327 nm², C = 0 ± 1 nm⁴). (d) the refractive index and its corresponding real part of the dielectric response of the $GaS_xO_y$.

To further investigate the structure and chemical composition of the oxide, transmission electron microscopy (TEM) has been employed in conjunction with electron energy loss spectroscopy (EELS) measurements. The TEM results provide an understanding of the material's atomic arrangements, while EELS maps provide the elemental distributions, allowing for a better understanding of the formed oxide nature. Figures 3 (a) and (b) show the cross-sectional TEM image of a partially oxidised GaS flake, subjected to 15-minute oxidation, distinctly showing the formation of oxide/semiconductor ($GaS_xO_y$/GaS) heterostructure with an oxide thickness of roughly 4 nm. The oxide derived in this manner is amorphous in nature, similar to oxides produced by oxygen plasma oxidation of other 2D materials[19,38]. Notably, the plasma oxidation does not damage the lattice of the unoxidised GaS which appears beneath the oxidised GaS. This is confirmed by in-plane TEM images of partially oxidised GaS flake, which were transferred on a Silicon Nitride ($SiN_x$) TEM grid before the oxidation. Here, the unoxidised GaS retains its inherent hexagonal crystalline structure, with the damage appearing minimal (Fig. 3(c)). This observation aligns well with the diffraction patterns presented in Fig. 3(d). In Fig. 3(e), we show the high-angle annular dark-field scanning TEM (HAADF-STEM) region from the same partial oxidised GaS flake as shown in Fig. 3(a,b). Figs. 3(f)-(h) presents the EELS elemental maps for Ga, S, and O corresponding to the imaged region in Fig. 3(e). These EELS results corroborate our earlier assertion that the partially oxidised GaS forms an oxide/semiconductor ($GaS_xO_y$/GaS) heterostructure, where the top layer undergoes oxidation while the bottom layer is minimally changed. However, the results also demonstrate that while there is a clear separation between the GaS and the $GaS_xO_y$ layers, the $GaS_xO_y$ still contains some sulfur species, as shown in Fig 3(g), and hence the designation $GaS_xO_y$.

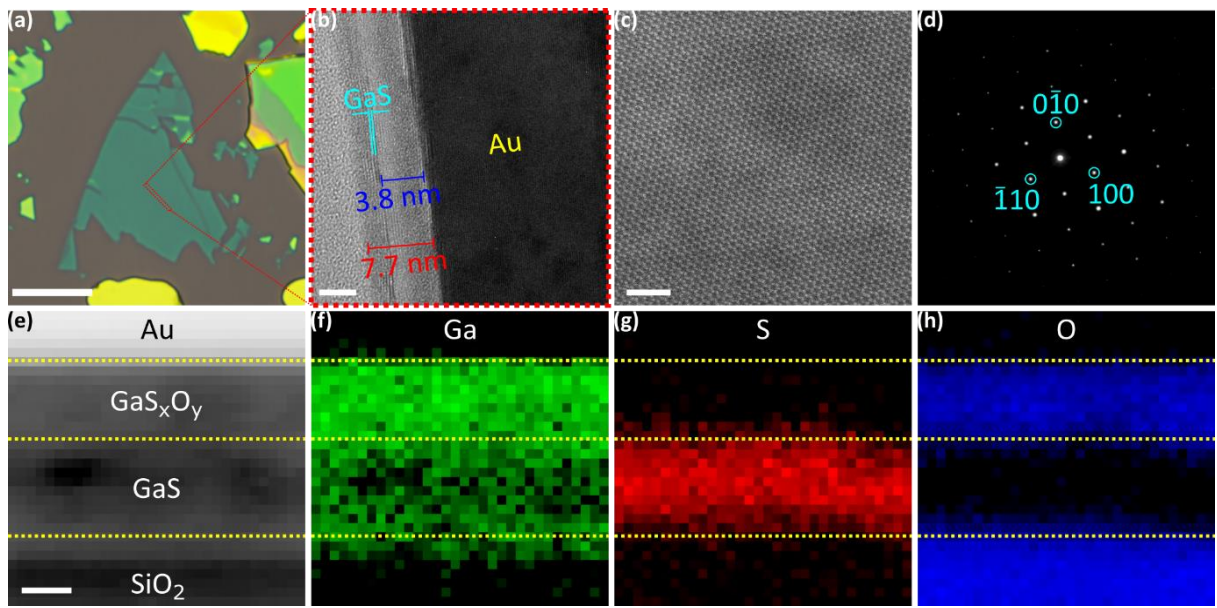

**Figure 3:** TEM images of partially oxidised GaS flake that was oxidised for 15 mins. (a) and (b) shows the optical image of the $GaS_xO_y$/GaS flake and its corresponding cross-section TEM image. The TEM image shows the formation of an amorphous layer on top of the crystalline GaS (scale 10 μm for optical and 3.8 nm for TEM). (c) and (d) HR-STEM image of a $GaS_xO_y$/GaS flake and its diffraction pattern confirming that the unoxidised GaS layers exhibit minimal damage (scale 2 nm). (e)-(h) HAADF-STEM image and its EELS elemental map of the flake from (a) showing the dominant sulfur and oxygen domain corresponding to the unoxidised GaS and the $GaS_xO_y$ layers, respectively (scale 2 nm).

As shown in Figs. 4(a) and (b), the TEM images of the $GaS_xO_y$ produced at different oxidation times indicate that the oxide thickness after 10 and 15 minutes of oxidation is almost indistinguishable, which could indicate a self-limiting mechanism for oxidation at the given oxidation parameters. The average oxide thickness for both oxidation times was found to be ~3.8 nm, with a standard deviation of ~0.37 nm (Fig. 4(c)). In the context of device fabrication, the oxidation technique demonstrated here presents a significant advantage by reliably producing ultra-thin oxide layers while exhibiting a wide tolerance for variations in oxidation times.

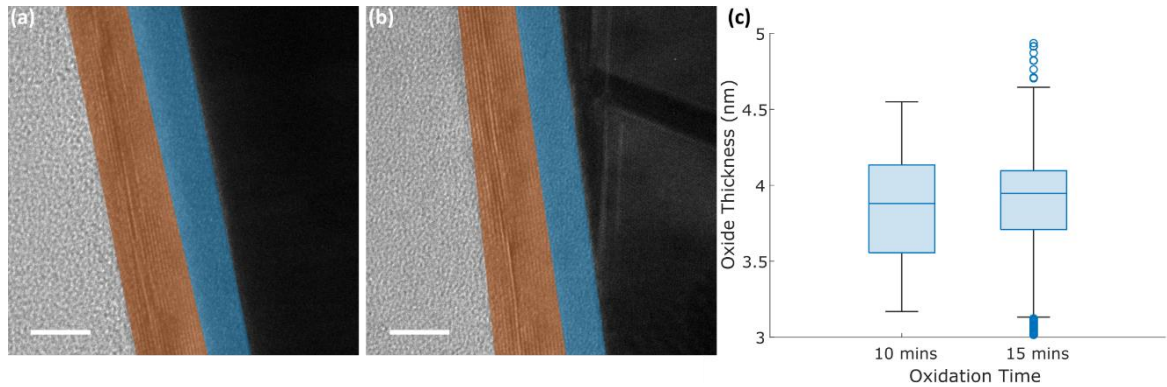

Figure 4: Time-dependent oxide thickness. (a) and (b) are false colour TEM cross-sectional images of 10 and 15 mins oxidation time, respectively. The GaS region was highlighted in orange, while the $GaS_xO_y$ region was highlighted in blue (scale 5 nm). (c) The calculated oxide thicknesses of 10 and 15 mins oxidation demonstrating near identical average and distribution (open circles are for thickness outliers, see Methods).

In the landscape of low-power electronics, the push for more efficient, compact, and faster memory technologies has been endless. Memristors stand out in this domain as promising candidates due to their unique ability to retain their resistance state based on applied voltage or current history[39]. At the forefront of this innovative frontier is the ReRAM, a distinct paradigm of switching memory technology. Beyond their evident advantages in terms of speed, endurance, and device density, ReRAM devices present a deceptively simple architecture[40-43]. This convergence of attributes makes ReRAMs key candidates for leading the advancement in low-power memory systems. Conventionally sputtered or deposited $Ga_2O_3$ has shown great potential in resistive switching applications with both uni- and bipolar behaviour observed[10-14,16]. In addition, recently, partially oxidised 2D materials such as $HfS_2$, $HfSe_2$, and $PdSe_2$ have shown promise for these applications[38,44-46], which motivated us to explore the use of the $GaS_xO_y$/GaS heterostructure in ReRAM applications.

As described in the methods section, the $GaS_xO_y$/GaS ReRAMs were fabricated using a 50 nm Au bottom electrode and a 90/10nm Au/Ti top electrode. Due to the small footprint of each device (~4 $\mu m^2$), multiple devices were made from the same flake. However, rather than using a standard crossbar array of multiple write lines (WL) and bit lines (BL), we used a single common BL to avoid sneak current paths from forming, as shown in Fig.5 (a). Figure 5 (b) shows typical DC I-V characteristics of GaS$_x$O$_y$/GaS ReRAM oxidised for 10 mins where the voltage bias is applied to the bottom electrode. The device exhibits a bipolar switching behaviour, switching from a high resistive state (HRS) to a low resistive state (LRS) with a negative DC I-V sweep and from LRS to HRS using a positive DC sweep. The low SET (V$_{SET}$) and RESET voltages (V$_{RESET}$) of -0.4 V and 0.3 V, respectively, can be attributed to several factors such as the ultra-thin (~4 nm) and low-density nature of the oxide in addition to the presence of mobile oxygen and sulfur ions. We note that all the 6 tested GaS$_x$O$_y$/GaS ReRAMs with Au/Ti top electrode start in LRS and, therefore, do not require a forming step, which is normally required for resistive memories[47-49]. This can be advantageous for resistive device applications as it eliminates the need for a separate circuit responsible for the forming step and avoids current overshoots. While the reason behind the forming-free behaviour is still unclear, it can be speculated that this could be a result of the lower density of the oxide and the potential access of mobile ions within the switching layer. To further investigate the GaS$_x$O$_y$/GaS ReRAM device performance, retention measurements were conducted. The device exhibits good retention at 0.2 V READ voltage, where we can see a clear separation of the resistive states for $10^4$ s (Fig.5 (c)). To characterise the device switching behaviour, pulsed voltage stress (PVS) measurements approach was chosen. The PVS measurements was performed by applying 500 ns, -0.5 V SET pulses and 2 μs, 0.5 V RESET pulses on a different device on the same flake. The resistance was read using 2 μs, 0.2 V pulses having a 100 ns rise and fall time and 2 μs separation time as shown in Fig. 5(d). As shown in Fig. 5(e), The GaS$_x$O$_y$/GaS ReRAM shows a slight variation in states over 350 cycles using PVS measurements with an average HRS/LRS ratio of ~3. In addition, the cumulative distribution function (CDF) plot, shown in Fig. 5(f), confirms the observation from the endurance measurement where a clear window exists between HRS and LRS with no overlap at any point. In order to assess the potential of our GaS$_x$O$_y$/GaS ReRAM for low-power applications, we calculated the energy consumed per SET/RESET operation by numerical integrating the power consumed within the pulse width used. The device consumes, on average, 0.22 nJ and 0.51 nJ to SET and RESET respectively. It is worth noting that the energy consumed by the RESET step excluding the plateau region is on average 0.31 nJ. While this exceeds the desired energy target for these applications, we believe that further optimisation of the oxide/semiconductor structure can yield lower energy consumption.

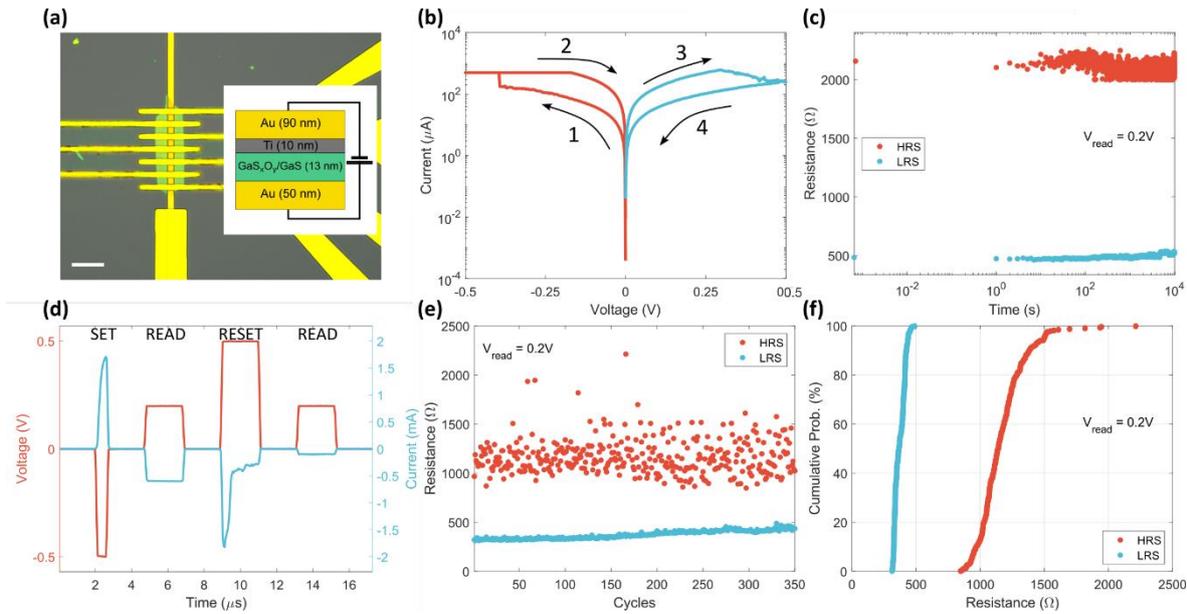

**Figure 5: Oxidised GaS-based ReRAM element electrical characterisation. (a)** Optical image of ReRAM element showing common bottom electrode with multiple top electrodes (scalebar: 10 μm). Inset is a schematic of the vertical stack ReRAM. **(b)** Typical DC switch cycle of ReRAM device where the device switches from HRS to LRS in the negative voltage regime and from LRS to HRS when voltage polarity changes. **(c)** resistance state retention over time measured at 0.2 V showing stable separation between HRS and LRS for $10^4$ s. **(d)** a representative single PVS SET and RESET measurement cycle of voltage and current showing a SET step followed by READ, and RESET step followed by another READ. **(e) and (f)** shows endurance measurement and corresponding CDF plots read at 0.2V.

Understanding the performance of ReRAM devices requires not only an examination of their electrical characteristics, but also a deeper insight into their switching mechanisms. Different device architectures exhibit varied level of complexity, which critically influence device performance. The fact that $GaS_xO_y$/GaS ReRAM has an additional interface which is the oxide/semiconductor one suggests that the switching mechanism is far more complex. In the case of Ti-based $Ga_2O_3$ resistive switching devices, the commonly suggested mechanism for switching is via the conventional oxygen vacancy filament [9,12,50]. However, in $GaS_xO_y$/GaS ReRAM, such a simple explanation might not be applicable. First, the oxide layer in $GaS_xO_y$ contains not only oxygen ions and their vacancies but also sulfur ions and their vacancies. While the oxygen ions have a smaller size compared to sulfur ions, which allows them to have higher mobility, determining the dominant conduction-driving ion depends on other factors, such as ions' concentration and the energy barrier for their movement within the oxide. However, a case can be made here for the oxygen ions based on the EELS results in Figs. 3(g) and (h). As the sulfur ions are more concentrated near the semiconductor while the oxygen ions are more abundant near the Ti top electrode, it is more likely to have the more mobile oxygen ions that are close to the Ti layer to contributing to the filament formation than the far and less mobile sulfur ions [38]. Nevertheless, it is also important to remember that the bottom semiconductor layer might have an active role in the switching. The role of the semiconductor layer is far more complex. While an oxide layer is needed to have switching, it is observed in literature that the presence of the semiconductor layer improves the stability and endurance of the resistive switching behaviour in the device [38,45].

The complexity of the $GaS_xO_y$/GaS ReRAM extends to the explanation of the conduction mechanism. Previous reports on the Ti-based $Ga_2O_3$ memristors suggest a space charge limited conduction (SCLC) mechanism in the HRS and ohmic conduction in the LRS [12,16,50,51]. This assertion seems to hold true for the $GaS_xO_y$/GaS ReRAM (tested on another device with 10 nm $GaS_xO_y$/GaS), as shown in Fig. 6(a). The HRS show the characteristics of Space Charge Limited Current (SCLC) domains with an average slope for the ohmic region of ~1.1, indicating linear current-voltage (I-V) behaviour typical of ohmic conduction and a Child regime slope of ~1.95 over 97 DC-IV cycles, which reflects a quadratic I-V relationship typical of the trap-filled limited conduction. However, the LRS state also shows two domains with an expected ohmic region with a ~1.01 slope but an additional ~1.57 slope region. This suggests the LRS states have at least two conduction mechanisms potentially as a result of the semiconductor layer Fig. 6(b).

Ti-based top electrode plays a significant role in helping facilitate the formation of a stable conductive filament, which allows the $GaS_xO_y$/GaS ReRAM to exhibit a resistive switching behaviour [52]. To investigate the role of the Ti layer, another set of devices was made without a Ti layer, i.e. consisting of a stack of Au/$GaS_xO_y$/GaS/Au as shown in Fig. 6(c). Devices based on this architecture failed to exhibit any stable bipolar switching behaviour. Initially, these devices may display one or two promising cycles with current compliance ($I_{CC}$) set at nanoampere levels but rapidly degrade to a state where no further switching behaviour can be observed. Hence, it can be concluded that the presence of Ti is crucial to facilitating the anionic filamentary formation [38]. Nevertheless, an unstable unipolar switching characteristic can be achieved under certain conditions, as shown in Fig. 6(c). It is unclear what drives such a switching mechanism in this type of device architecture. However, based on previous reports, an Au metallic filament could be responsible for the switching in memristors with only Au contacts [53,54].

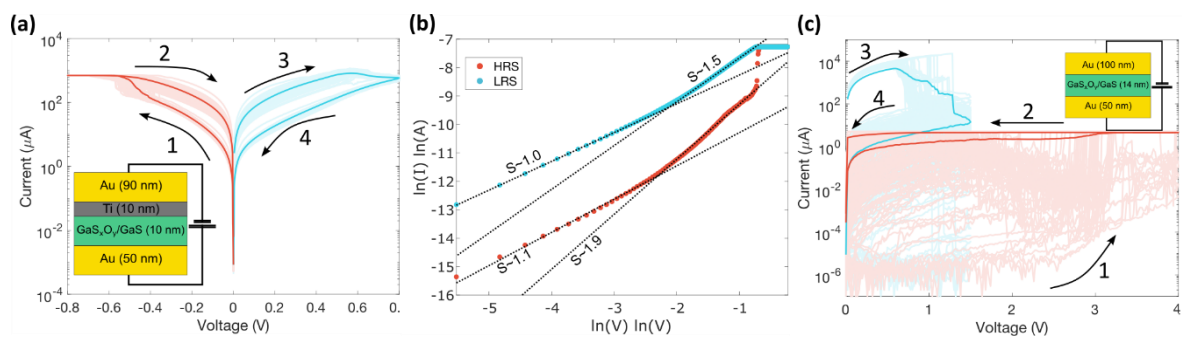

**Figure 6: Electrode role in device switching in $GaS_xO_y$/GaS ReRAM.** (a) DC IV sweep of the devices with light colours representing the cycles and the saturated colour curves representing the average. Inset is a schematic of the vertical stack ReRAM with 10 nm $GaS_xO_y$/GaS. (b) ln(I) vs ln(V) plot of a representative IV curve showing the characteristic SCLC domain

in HRS while also showing different domains in the LRS. (c) Unipolar DC switch cycle of Au only electrode GaS$_x$O$_y$/GaS ReRAM where light colours representing the cycles and the saturated colour curves representing the average. Inset is a schematic of the Au only vertical stack ReRAM.

## Conclusions

In this study, we have successfully demonstrated the capacity to oxidise GaS, a representative of the group III monochalcogenides, via a controlled oxygen plasma treatment. Upon close inspection, it was discovered that the oxidation process starts at the surface and progressively moves inward. The resultant oxide is amorphous with a low density and contains some sulfur species doping. Intriguingly, this imperfect oxide composition has proven to be advantageous for ReRAM applications. Our evaluations of ReRAM devices that incorporated this oxide have shown significant potential combining forming-free operation with hours long retention, endurance of hundreds of cycles, and sub nano-Joule energy consumption. With further optimization of the oxidation and careful design of oxide-semiconductor we expect that such devices can play a significant role in low power electronics. Moreover, the oxidation method can be explored for applications beyond ReRAMs such as UV photodetectors and gas sensing.

## Acknowledgements

AA acknowledges the financial support of the Saudi Arabian ministry of education. AX and AL acknowledge funding from EPSRC grant EP/T517793/1. VN and XG wish to thank the support of the Science Foundation Ireland funded AMBER research centre (Grant No. 12/RC/2278_P2), and the Frontiers for the Future award (Grant No. 20/FFP-A/8950). Furthermore, VN and XG wish to thank the Advanced Microscopy Laboratory in CRANN for the provision of their facilities. S.H. and I.C. acknowledge funding from EPSRC (EP/T001038/1, EP/V047515/1, EP/N509620/1, EP/R513180/1).

# Methods

*GaS samples preparation:*

GaS flakes were prepared by micromechanical exfoliation from bulk GaS crystal (HQgraphene). $SiO_2$/Si substrates were first cleaned using acetone, followed by isopropyl alcohol (IPA) and oxygen plasma treatment using an Advanced Vacuum–vision 320 RIE. The substrates were exposed to 100 W $O_2$ plasma at 40 sccm and 20 mTorr for 2 mins and then transferred into a glovebox with an Ar environment and $O_2$/$H_2O$ level <0.1 ppm. The flakes were exfoliated immediately on the pre-treated substrate. Flakes of interest were identified using an optical microscope. Both exfoliation and optical microscope mapping of flakes were done in the same Ar environment.

*GaS plasma oxidation:*

Advanced Vacuum – vision 320 RIE was used to perform the $O_2$ plasma oxidation. Before placing the substrate, a 100 W $O_2$ plasma cleaning step was run at 40 sccm and 20 mTorr for 15 mins to clean the chamber. To reduce the effect of ambient oxidation, the samples were maintained inside the glovebox until the oxidation chamber was ready. The samples were exposed to ambient air for less than 30s. For oxidation, 10 W $O_2$ plasma was used at 10 sccm and 20 mTorr for 10 and 15 mins. Post oxidation, the samples were moved back immediately to the glovebox.

*Device fabrication:*

For device fabrication, a high resistivity (>10,000 Ω cm) Si substrate with 285 nm $SiO_2$ was used to prevent any unwanted current paths during pulse operations due to parasitic capacitances. Substrates were cleaned with acetone and IPA, followed by 100 W $O_2$ plasma at 40 sccm and 20 mTorr for 2 mins. Contacts were patterned by maskless photolithography utilising a Microtech LW405 laser writer. AZ-5214E image reversal photoresist was chosen for this purpose due to the ease of the lift-off process. Using Kurt J. Lesker PVD 200 Pro, a 50 nm Au bottom electrode was deposited. For lift-off, the substrates were left in an acetone bath overnight. GaS flakes were exfoliated on Si/$SiO_2$ substrate as described previously. Here, plasma treatment was skipped as it reduces the success rate of flake transfer. Selected flakes, identified by using an optical microscope, were transferred on top of the bottom electrode using a conventional dry transfer technique via Polydimethylsiloxane/polycarbonate (PDMS/PC) stamp[55]. Chloroform was used to remove any residual PC, followed by an IPA cleaning step. While the transfer was done in a glovebox, the chloroform step was done in ambient conditions. Extra care was taken to reduce sample exposure to oxygen during this process. Once transfer and cleaning were completed, the GaS flakes were oxidised as described above. The top electrode was fabricated following the same procedure described for the bottom electrode with the difference that electrode consisted of 10nm Ti and 90nm of Au. Ti was deposited under ~1x10$^{-7}$ mTorr pressure to prevent Ti oxidation. For the device shown in Fig, 5, only Au was used.

*Electrical Measurements:*

Keysight/Agilent B1500A semiconductor device parameter analyser was utilised for all electrical measurements. Voltage was always applied through the common bottom electrode, whereas the top electrode was grounded. A Keysight B1530A Waveform Generator/Fast Measurement Unit (WGFMU) module was used for pulse measurements in conjunction with a Remote-sense and Switch Unit (RSU). All measurements were done in ambient conditions at room temperature.

*Raman Spectroscopy Measurements:*

Raman measurements were performed using a Renishaw inVia microspectrometer. A 514.5 nm laser was used for all measurements with a 100X magnification lens and a 2,400 line/mm grating. The time and power during the Raman measurement were minimised to avoid unintentional photo-oxidation.

*AFM Measurements:*

The AFM measurements were conducted on a Bruker Dimension Icon AFM. Peak Force Tapping mode coupled with ScanAsyst algorithm was used for all the measurement.

*Ellipsometry Measurements:*

We conduct ellipsometry measurements using an EP4 Spectroscopic Imaging Ellipsometer from Park Systems Gmbh. The equipment is equipped with a laser-stabilized Xenon arc lamp and a three-grating monochromator as the spectroscopic light source. This allows for a spectral range between 360 nm to 1000 nm, with a 5 nm bandwidth of the output line. We use a polarisation state generator (PSG), consisting of a linear polariser and quarter-wave plate (compensator), to control the state of polarisation of the incident beam. The reflected light is collected through an analyser and a 50x objective to a 1392x1040 pixel CCD camera, which allows for a lateral resolution down to 1 μm. We perform all measurements in the P-A-nulling mode, where the compensator is kept fixed at 45°, and the polariser and analyser positions are shifted until the signal minimum is reached. Once the nulling condition is satisfied, we record the sample's ellipsometric angles Δ and ψ at each probing wavelength.

*TEM Measurements:*

To prepare samples for cross-sectional TEM measurement, we exfoliated GaS flakes on SiO$_2$/Si substrate as described above. The flakes were then oxidised to the desired time. To provide protection during the lamella preparation process, a double layer consisting of Au and aluminium oxide (AlO$_x$) was deposited using e-beam evaporation and ALD respectively. During the ALD step, extra care was taken to minimise temperature (which was kept below 150 °C) and time to prevent any modification of the oxide. Lamellas were fabricated using a Dual Beam FIB-SEM system (Carl Zeiss Auriga) equipped with a platinum (Pt) deposition cartridge and nanomanipulator (Kleindiek Nanotechnik). The process started by depositing ~100 nm of platinum (Pt) as an additional protection deposited by e-beam evaporation (2 kV). This was subsequently overlaid with a micrometre-thick platinum layer deposited via a gallium ion beam. After the preliminary rough milling stage at 30 kV, the resulting plate, approximately 2 µm in thickness, was extracted and attached on the edge of a copper finger in a FIB lift-out grid. The lamella was subsequently thinned using a 15 kV ion beam to minimise ion beam-induced damage, resulting in a final lamella with a thickness under 100 nm. After thinning, the lamella was transferred to the TEM column for observation as soon as possible to minimise potential further oxidation.

The in-plane samples for (S)TEM were prepared by exfoliating GaS flakes as described above. The chosen flakes were then dry transferred using the PDMS/PC transfer method to silicon nitride ($SiN_x$) TEM grid with 1 µm diameter holes. Chloroform was used to remove PC residual. The transferred flake was then plasma oxidised as described above.

Transmission electron microscopy (TEM) and scanning TEM (STEM) was conducted on an uncorrected FEI Titan equipped with a Schottky field emission S-FEG source operating at 300 kV. Electron energy-loss spectroscopy (EELS) mapping was executed employing a Quantum Gatan Imaging Filter (GIF) detector, possessing an energy dispersion of 0.5 eV per channel.

To measure the oxide thickness from the TEM images, two flakes per oxidation time were used. To ensure that the TEM images are representative of the flake, at least 5 TEM cross-sectional images were taken per flake at different points. All 20 images were taken using the same magnification to ensure consistency. The images were then processed using purposely made MATLAB code to calculate the oxide thickness in each image. It should be noted that the code requires the user to define the boundaries of each region. To differentiate between the oxide and semiconductor layers, the user manually selects at least 10 points at the interface of the layers, and the code employs interpolation to create a detailed 100-point trace along the edge. The thickness measurement is subsequently based on the analysis of these 100 points, ensuring precise delineation between the amorphous oxide layer and the semiconductor.